\documentclass[12pt]{article}%
\usepackage{amsmath}
\usepackage{amsfonts}
\usepackage{amssymb}
\usepackage{graphicx}%
\providecommand{\U}[1]{\protect\rule{.1in}{.1in}}
\begin{document}

\begin{center}
{\LARGE \textbf{Contest vs. Competition in} \textbf{Cournot\smallskip}}

{\LARGE \textbf{Duopoly: Schaffer's Paradox}}{\large \footnote{The authors are
grateful to Thorsten Hens, Alexander Matros, and Larry Samuelson for helpful
discussions of questions related to unbeatable strategies and their
applications.}}{\huge \bigskip}{\LARGE \textbf{\smallskip}}

R. Amir\footnote{Department of Economics, University of Iowa, Iowa City, IA
52242-1994, USA.{} E-mail: rabah-amir@uiowa.edu.}, I. V.
Evstigneev\footnote{Department of Economics, University of Manchester, Oxford
Road, Manchester, M13 9PL, UK.\ E-mail:: igor.evstigneev@manchester.ac.uk.},
and M. V. Zhitlukhin\footnote{Steklov Mathematical Institute, Russian Academy
of Sciences, Moscow, 119991, Russia. E-mail: mikhailzh@mi-ras.ru.
(Corresponding Author.)}{\huge \bigskip}
\end{center}

\begin{quotation}
\textbf{Abstract. }The paper compares two types of industrial organization in
the Cournot duopoly: (a) the classical one, where the market players maximize
profits and the outcome of the game is a Cournot-Nash equilibrium; (b) a
contest in which players strive to win a fixed prize/bonus employing
unbeatable strategies. Passing from (a) to (b) leads to a perfect competition
with zero profits of the players (Schaffer's paradox). Transition from (b) to
(a) results in a substantial decline in the production output, which also
seems paradoxical, as it is commonly accepted that competition increases
efficiency. We examine these phenomena in two versions of the Cournot model:
with a homogeneous good and with differentiated goods. \bigskip

\textbf{Keywords} \ Cournot duopoly, Borel's unbeatable strategies, Relative
preferences, Evolutionary game theory, Schaffer's paradox

\textbf{Mathematics Subject Classification -- MSC2020:} 91B24, 91B54, 91A22 

\smallskip
\end{quotation}

\section{Introduction}

\textbf{1.1. Cournot contest. }A firm producing a homogeneous good owns two
production units (plants). Plants $i=1,2$ are run by two managers modeled as
players in the Cournot duopoly game with inverse demand $P(Q)$ and profits
\begin{equation}
\pi_{i}(q_{1},q_{2})=q_{i}P(Q)-cq_{i}\ (i=1,2,\ Q=q_{1}+q_{2}),
\label{homogeneous}%
\end{equation}
where $c>0$ is the production cost, the same for both plants. The players
select their strategies (the quantities they produce) $q_{i}\geq0$
independently and simultaneously. The goal of the firm, serving the whole
market, is to maximize profits. To achieve this goal it contemplates an
incentive scheme for the managers. The standard way of doing it would be to
share with the managers a fixed percentage of profits. This would result in
the classical Cournot-Nash equilibrium in the game at hand.

However, the "parsimonious" firm, rather than sharing with the managers some
fixed percentage of profits, decides to set a contest. A fixed prize/bonus
$B>0$ is awarded to that manager who succeeds in getting a higher profit than
the other. If they get equal profits, they share the award equally: each gets
$B/2$.

Suppose there exists a strategy $q^{\ast}$ that would guarantee a positive
bonus ($B$ or $B/2$) for player 1 irrespective of the strategy of player 2.
Then $q^{\ast}$ must satisfy%
\begin{equation}
\pi_{1}(q^{\ast},q)\geq\pi_{2}(q^{\ast},q), \label{unb}%
\end{equation}
for all $q\geq0$. Strategies of this kind are called \textit{unbeatable}. They
allow player 1 to outperform the rival (or at least to achieve the same
result) in terms of profit maximization, and hence in terms of the
reward/bonus, irrespective of the rival's strategy. Since the game at hand is
symmetric, the sets of unbeatable strategies for players 1 and 2 coincide.

The idea of unbeatable strategies goes back to Borel, who wrote
(\cite{Borel1921}, p. 1304):

\begin{quote}
{\footnotesize One may propose to investigate whether it is possible to
determine a method of play better than all others; i.e., one that gives the
player who adopts it a superiority over every player who does not adopt it.}
\end{quote}

\noindent This idea served as a starting point for von Neumann's
\cite{VonNeumann1928} seminal work on zero-sum games, motivated primarily by
the economic applications. In quite a different context, the concept of an
unbeatable strategy re-emerged several decades later in evolutionary biology
(Hamilton \cite{Hamilton1967}). It served as a germ for the notion of an
evolutionary stable strategy, which became central to evolutionary game theory
(Maynard Smith and Price \cite{MaynardSmithPrice1973}, Maynard Smith
\cite{MaynardSmith1982}). Kojima \cite{Kojima2006} was the first to apply the
evolutionary theory of unbeatable strategies in economics. Different lines of
studies related to unbeatable strategies have been synthesized in the paper by
Amir at al. \cite{AmirEvstigneevPotapova2024}, presenting the subject in a
modern perspective.

If a strategy $q^{\ast}$ with property (\ref{unb}) exists, it will be rational
for the participants of the contest to select it. Indeed, $q^{\ast}$ will
guarantee a bonus of at least $B/2$, whatever the rival undertakes. If some
strategy of the first player is not unbeatable, then the second player can act
so as to get a strictly higher profit than the first, in which case the first
one will get no bonus. Thus those and only those strategies represent
\textit{solutions to the contest} that are unbeatable.

\textbf{1.2. Schaffer's paradox. }Assume that the inverse demand function
$P(Q)$ satisfies the following condition:

(C) There exists a quantity level $\bar{Q}>0$ such that (i) $P(\bar{Q})=c$,
(ii) $P(Q)<c$ for each $Q>\bar{Q}$, and (iii) $P(Q)>c$ for each $Q<\bar{Q}$ .

\textbf{Proposition 1.1. }\textit{An unbeatable strategy }$q^{\ast}%
$\textit{\ solving the Cournot contest exists, is unique, the same for both
players, and is given by }$q^{\ast}=\bar{Q}/2$\textit{. If both players use
}$q^{\ast}$\textit{, then the outcome of the game is as follows: the total
production output is equal to} $\bar{Q}$,\textit{\ the commodity price
coincides with the production cost, and the profits of both production units
are equal to zero:}%
\[
2q^{\ast}=\bar{Q},\ P(\bar{Q})=c,\ \pi_{i}(q^{\ast},q^{\ast})=0.
\]

Clearly, this paradoxical outcome is disastrous for the profit maximizing
firm. It wildly contradicts the original goal of the contest designed to
create a cost-efficient incentive scheme for profit maximization. The result
contained in Proposition 1.1 is inspired by the seminal work of Schaffer
(1988, 1989), and we refer to it as \textit{Schaffer's paradox}.

Note that condition (C) holds (with $\bar{Q}=1-c$) for the linear inverse
demand $P(Q)=1-Q$, and then we have
\[
q^{\ast}=(1-c)/2,\ \bar{Q}=1-c,\ P(\bar{Q})=c,\ \pi_{i}(q^{\ast},q^{\ast})=0.
\]

\textbf{1.3. Focus of the paper. }In this work we perform a comparative
analysis of the outcomes of contest and Nash equilibrium in Cournot duopoly
models with a homogeneous good (\ref{homogeneous}) and differentiated goods,
see (\ref{differentiated}) and (\ref{fi}). The latter model is described as
follows. A firm owns two production units/plants supplying to the market two
different goods, each of which is to a certain extent a substitute to the
other. Firm 1 produces the quantity $q_{1}\geq0$ of good 1, and firm 2
produces the quantity $q_{2}\geq0$ of good 2. The production decisions
(strategies) $q_{1}$ and $q_{2}$ are selected by the managers of the plants
simultaneously and independently of each other. The inverse demand functions,
specifying, the market clearing prices, for the two goods are
\begin{equation}
P_{1}(q_{1},q_{2})=1-q_{1}-bq_{2},\;P_{2}(q_{1},q_{2})=1-q_{2}-bq_{1}.
\label{differentiated}%
\end{equation}
The number $b\in(0,1]$ reflects the extent to which the firms' products
substitute each other. The payoff functions (profits) are given by
\begin{equation}
\phi_{i}(q_{1},q_{2})=q_{i}P_{i}(q_{1},q_{2})-cq_{i},\;i=1,2\ , \label{fi}%
\end{equation}
where $0<c<1$ is the marginal cost of production. The case $b=1$ corresponds
to the Cournot model with a homogeneous good and linear inverse demand.

We consider a contest analogous to that described in \textbf{1.1. }A fixed
prize (bonus) $B>0$ is awarded to that manager whose production unit gets a
higher profit than the other. In the case of equal profits, each gets $B/2$.
The main results are as follows. We find the classical Cournot-Nash
equilibrium and the unbeatable strategy, the same for both players, solving
the corresponding contest. We compare them and examine their dependence on the
coefficient of substitutability $b$. Special attention is paid to the analysis
of a phenomenon similar to Schaffer's paradox. Remarkably, its paradoxical
features become "milder" when $b>0$ decreases, taking on their extreme forms
in the case of a homogeneous good ($b=1$).

The paper is organized as follows. Section 2, focusing on the case of a
homogeneous good, gives a proof of Proposition 1.1. Section 3 examines the
model with differentiated goods. Section 4 conducts a comparative analysis of
the outcomes of the contest and competition in the Cournot duopoly, paying
special attention to their dependence on the coefficient of substitutability
$b$. Section 5 concludes.

\section{Homogeneous good}

\textbf{2.1. }\textit{Proof of Proposition 1.1.} We will show that%
\begin{equation}
\pi_{1}(q^{\ast},q)>\pi_{2}(q^{\ast},q)\ \ \text{for }q\neq q^{\ast}.
\label{str-unb}%
\end{equation}
This means that the strategy $q^{\ast}$ is not only unbeatable, but strictly
outperforms any strategy $q$ distinct from $q^{\ast}$ (i.e. it is
\textit{strictly unbeatable}).

Observe that when both players use $q^{\ast}$, the total output will be
$\bar{Q}/2+\bar{Q}/2=\bar{Q}$. Therefore $P(\bar{Q})=c$ by virtue of
assumption (C), and so the profit of each player is zero:%
\[
\pi_{i}(q^{\ast},q^{\ast})=q_{i}P(\bar{Q})-cq_{i}=cq_{i}-cq_{i}=0.
\]

Let us prove (\ref{str-unb}). Put $f(q_{1},q_{2})=\pi_{1}(q_{1},q_{2})-\pi
_{2}(q_{1},q_{2})$. Then we have%
\[
f(q_{1},q_{2})=q_{1}P(q_{1}+q_{2})-cq_{1}-q_{2}P(q_{1}+q_{2})+cq_{2}=
\]
\[
(q_{1}-q_{2})P(q_{1}+q_{2})-c(q_{1}-q_{2})=[P(q_{1}+q_{2})-c](q_{1}-q_{2}).
\]
If $q>q^{\ast}$, then $q^{\ast}+q>\bar{Q}$, and by virtue of (C), $P(q^{\ast
}+q)-c<0$. Since $q^{\ast}-q<0$, we have
\[
f(q^{\ast},q)=[P(q^{\ast}+q)-c](q^{\ast}-q)>0.
\]
If $q<q^{\ast}$, then $q^{\ast}+q<\bar{Q}$, and according to (C), $P(q^{\ast
}+q)-c>0$. Therefore $f(q^{\ast},q)>0$. Finally, if $q=q^{\ast}$, then
$f(q^{\ast},q)=0$. Consequently, (\ref{str-unb}) holds, i.e., $q^{\ast}%
=\bar{Q}/2$ is a strictly unbeatable strategy.

Let us prove that $q^{\ast}=\bar{Q}/2$ is unique: if $q^{\prime}$ is an
unbeatable strategy, then $q^{\prime}=\bar{Q}/2$. Suppose $q^{\prime}\neq
\bar{Q}/2$ is an unbeatable strategy. Put $q=\bar{Q}/2$. If $q^{\prime}%
<\bar{Q}/2$,\ then $q^{\prime}-q<0$ and $q^{\prime}+q<\bar{Q}$, which yields
$P(q^{\prime}+q)-c>0$ and so $f(q^{\prime},q)=(q^{\prime}-q)[P(q^{\prime
}+q)-c]<0$. This contradicts the assumption that $q^{\prime}$ is an unbeatable
strategy. If $q^{\prime}>\bar{Q}/2$,\ then $q^{\prime}-q>0$ and $q^{\prime
}+q>\bar{Q}$, which yields $P(q^{\prime}+q)-c<0$ and so $f(q^{\prime
},q)=(q^{\prime}-q)[P(q^{\prime}+q)-c]<0$. A contradiction. Thus, the
assumption that there exists an unbeatable strategy $q^{\prime}$ distinct from
$\bar{Q}/2$ in all the above cases leads to a contradiction. $\square$

\section{Differentiated goods}

\textbf{3.1. Nash equilibrium.} The goal of this section is to examine
phenomena similar to those we considered in the previous section, but for a
different version of the Cournot model: Cournot duopoly with differentiated
goods and linear inverse demand, as described in (\ref{differentiated}) and
(\ref{fi}).

In Proposition 3.1 we find the Nash equilibrium in the game at hand (cf. Singh
and Vives \cite{SinghVives1984}). Assuming that the players use the
equilibrium strategies, we compute the corresponding quantities $\hat{q}_{1}$
and $\hat{q}_{2}$, the total output $\hat{Q}=\hat{q}_{1}+\hat{q}_{2}$, the
prices $\hat{p}_{i}=P_{i}(\hat{q}_{1},\hat{q}_{2}$) of goods $i=1,2$, and the
profits $\hat{\phi}_{i}=\phi_{i}(\hat{q}_{1},\hat{q}_{2})$ of players $i=1,2$.

\textbf{Proposition 3.1. }\textit{We have}%
\begin{equation}
\hat{q}:=\hat{q}_{1}=\hat{q}_{2}=\frac{1-c}{2+b},\ \hat{Q}=\frac{2(1-c)}{2+b},
\label{q-Q}%
\end{equation}%
\begin{equation}
\hat{p}:=\hat{p}_{1}=\hat{p}_{2}=\frac{1+c(1+b)}{2+b},\ \hat{\phi}_{i}%
=\frac{(1-c)^{2}}{(2+b)^{2}}. \label{p-fi}%
\end{equation}

\textit{Proof. }Let us find the Nash equilibrium $(\hat{q}_{1},\hat{q}_{2})$.
The quantity $\hat{q}_{1}$ has to maximize%
\[
q_{1}(1-q_{1}-b\hat{q}_{2})-cq_{1}\ \text{over }q_{1}\geq0\text{.}%
\]
The first order optimality condition
\[
-2\hat{q}_{1}+1-c-b\hat{q}_{2}=0
\]
gives $\hat{q}_{1}=(1-c-b\hat{q}_{2})/2$. Analogously, for the production unit
2, we find $\hat{q}_{2}=(1-c-b\hat{q}_{1})/2$. By solving the system of
equations%
\[
\hat{q}_{1}=(1-c-b\hat{q}_{2})/2,\ \hat{q}_{2}=(1-c-b\hat{q}_{1})/2,
\]
we obtain that the $\hat{q}:=\hat{q}_{1}=\hat{q}_{2}$ satisfies%
\[
\hat{q}=(1-c-b\hat{q})/2.
\]
Thus%
\[
\hat{q}=\hat{q}_{1}=\hat{q}_{2}=\frac{1-c}{2+b},
\]
which proves (\ref{q-Q}).

To compute the equilibrium prices $\hat{p}_{i}=P_{i}(\hat{q}_{1},\hat{q}_{2})$
we write%
\[
\hat{p}_{i}=P_{i}(\hat{q}_{1},\hat{q}_{2})=1-\hat{q}-b\hat{q}=1-\frac
{1-c}{2+b}(1+b)=\frac{1+c(1+b)}{2+b}.
\]
The equilibrium profits are computed as follows:%
\[
\phi_{i}(\hat{q},\hat{q})=\hat{q}[P_{i}(\hat{q},\hat{q})-c]=\hat{q}%
(\frac{1+c(1+b)}{2+b}-c)=\hat{q}\frac{1-c}{2+b}=\frac{(1-c)^{2}}{(2+b)^{2}}.
\]
\ Thus, all the formulas in (\ref{q-Q}) and (\ref{p-fi}) are verified.\hfill
$\square$

In Proposition 3.2 below, we consider the contest described in \textbf{1.3}
for the Cournot duopoly with differentiated goods. We find the unbeatable
strategy $q^{\ast}$ solving the contest (it is the same for both players by
symmetry). Assuming that both players use $q^{\ast}$, we compute the total
output $Q^{\ast}=2q^{\ast}$, the prices $p_{i}^{\ast}=P_{i}(q^{\ast},q^{\ast}%
$) of goods $i=1,2$, and the profits $\phi_{i}^{\ast}=P_{i}(q^{\ast},q^{\ast
})$ of players $i=1,2$.

\textbf{Proposition 3.2.} \textit{The unbeatable strategy }$q^{\ast}$\textit{
solving the contest under consideration is given by}%
\begin{equation}
q^{\ast}=\frac{1-c}{2},\ Q^{\ast}=1-c, \label{contest-unb}%
\end{equation}
\textit{Further, we have}%
\begin{equation}
p^{\ast}:=p_{1}^{\ast}=p_{2}^{\ast}=\frac{1+c}{2}-\frac{b(1-c)}{2}%
,\ \phi^{\ast}:=\phi_{1}^{\ast}=\phi_{2}=\frac{(1-c)^{2}(1-b)}{4}.
\label{contest-Q}%
\end{equation}

Note that the unbeatable strategy $q^{\ast}$ and the total output $Q^{\ast}$
do not depend on $b$!

\textit{Proof of Proposition 3.2. }By definition, an unbeatable strategy
$q_{1}=q^{\ast}$ of player 1 satisfies%
\[
q^{\ast}(1-q^{\ast}-bq)-cq^{\ast}\geq q(1-q-bq^{\ast})-cq,
\]
or equivalently,%
\[
q^{\ast}(1-q^{\ast})-cq^{\ast}\geq q(1-q)-cq.
\]
Thus $q^{\ast}$ maximizes%
\[
q(1-q)-cq=-q^{2}+q(1-c).
\]
Consequently, the unbeatable strategy $q^{\ast}$ of player 1 (and by symmetry,
of player 2) prescribes to produce the quantity (\ref{contest-unb}), which
yields the total output (\ref{contest-Q}).

To compute the prices resulting from the contest we write%
\[
p_{i}^{\ast}=P_{i}(q^{\ast},q^{\ast})=1-q^{\ast}-bq^{\ast}=1-(1+b)q^{\ast}%
\]%
\[
=1-(1+b)(1-c)/2=\frac{1+c}{2}-\frac{b(1-c)}{2}.
\]
Finally, we get\
\[
\phi_{i}^{\ast}=q^{\ast}(p_{i}^{\ast}-c)=\frac{1-c}{2}\cdot\lbrack\frac
{1+c}{2}-\frac{b(1-c)}{2}-c]=\frac{(1-c)^{2}(1-b)}{4}.
\]

\hfill$\square$

\section{Comparative analysis}

\textbf{Proposition 4.1. }\textit{Total equilibrium output}
\[
2\hat{q}=\frac{2(1-c)}{(2+b)}%
\]
\textit{is always not greater than the total contest output}
\begin{equation}
2q^{\ast}=1-c, \label{cont-output-homog}%
\end{equation}
\textit{when both players employ unbeatable strategies. }

\textit{Proof:} straightforward.\hfill$\square$

Note that in the case of a homogeneous good ($b=1$), we have%
\begin{equation}
2\hat{q}=\frac{2(1-c)}{2+b}=\frac{2(1-c)}{3}. \label{equil-output-homog}%
\end{equation}
Thus, when passing from contest to competition we observe, paradoxically, a
decline in the production output ("transformational recession", cf.
\cite{Kornai1994}). Why paradoxically? Because there is a common perception
that competition increases efficiency.

Note that the depth of the recession depends on the degree of substitutability
$b$ of the goods produced by plants 1 and 2. It increases when $b$ increases.
The maximum depth of the recession is observed when $b=1$, in the case of a
homogeneous good. In this case, production falls by $1/3$, see the last two formulas).

\textbf{Proposition 4.2. }\textit{The equilibrium profit}%
\[
\hat{\phi}=\frac{(1-c)^{2}}{(2+b)^{2}}%
\]
\textit{is always not less than the contest profit}
\[
\phi^{\ast}=\frac{(1-c)^{2}(1-b)}{4},
\]
\textit{as long as both players employ unbeatable strategies.}

\textit{Proof.} Indeed, the inequality $\hat{\phi}\geq\phi^{\ast}$ is
equivalent to $(2+b)^{2}(1-b)\leq4$. The last relation is true because the
derivative $\gamma^{\prime}(b)$ of the function $\gamma(b):=(2+b)^{2}(1-b)$,
which is equal to $-6b-3b^{2}$,
\[
\gamma^{\prime}(b)=2(2+b)(1-b)-(2+b)^{2}=4+2b-4b-2b^{2}-4-4b-b^{2}=
\]
is negative on $[0,1]$, and $\gamma(0)=4$.\hfill$\square$

It is important to note that the outcome of the contest in the case of a
homogeneous good (i.e. when $b=1$), is that of \textit{perfect competition:}
the profit $\phi^{\ast}$ is equal to zero. In this case, in the course of
transition from contest to competition, the profit increases from zero to
$\hat{\phi}=(1-c)^{2}/9$. If $0<b<1$, the profit increases as well, but not
that drastically. In the extreme case $b=0$, we have $\hat{\phi}=\phi^{\ast
}=(1-c)^{2}/4$.

\section{Concluding remarks}

To conclude we would like to outline some prospective topics for further research.

1) Comparative analysis of unbeatable strategies and Nash equilibrium in
various classical games, in particular those pertaining to industrial organization.

2) Investigation of unbeatable strategies in dynamic settings, in particular,
Stackelberg ones. Comparison of the outcomes in static and dynamic versions of
the game.

3) The study of two-stage games where the players first employ unbeatable
strategies and then conclude the game in a Nash equilibrium framework.

4) Reflecting in the models possibilities of a spiteful behaviour aimed at
increasing the relative performance. This is a classical subject in
evolutionary biology (see \cite{Hamilton1970}), potentially admitting
translation into the language of economics.

5) It would be of interest to study unbeatable strategies in asymmetric
models, where they are typically non-unique (if they exist). Then additional
criteria come into play, e.g., to outperform the rival to the greatest extent,
or to achieve this goal by using the smallest amount of resources. Some
results in this direction are obtained in \cite{AmirEvstigneevPotapova2024},
Sect. 5.

\end{document}